\documentclass[preprint,12pt]{elsarticle}

\usepackage{amssymb}
\usepackage{amsmath}

\usepackage{hyperref}
\usepackage{algorithm}
\usepackage{algpseudocode}
\usepackage{subcaption}
\usepackage{array}
\usepackage{xurl}
\usepackage{rotating}
\usepackage{amssymb}
\usepackage{pifont}
\newcommand{\cmark}{\textcolor{teal}{\ding{51}}}
\newcommand{\xmark}{\textcolor{red}{\ding{55}}} 

\usepackage{tikz}
\usepackage{booktabs}

\usepackage[most]{tcolorbox}
\tcbset{
  colback=gray!10,
  colframe=black,
  boxrule=1.5pt,
  arc=2mm,
  left=3mm,
  right=3mm,
  top=3mm,
  bottom=3mm,
  fonttitle=\bfseries
}

\begin{document}

\begin{frontmatter}



\title{
AURA: A Multi-Agent Intelligence Framework for Knowledge-Enhanced Cyber Threat Attribution
} 


\author[label1]{Nanda Rani} 
\author[label1]{Sandeep Kumar Shukla} 

\affiliation[label1]{organization={Department of Computer Science and Engineering},
            addressline={\\Indian Institute of Technology Kanpur}, 
            country={India}}

\begin{abstract}
Effective attribution of Advanced Persistent Threats (APTs) increasingly hinges on the ability to correlate behavioral patterns and reason over complex, varied threat intelligence artifacts.
We present AURA (Attribution Using Retrieval-Augmented Agents), a multi-agent, knowledge-enhanced framework for automated and interpretable APT attribution. AURA ingests diverse threat data including Tactics, Techniques, and Procedures (TTPs), Indicators of Compromise (IoCs), malware details, adversarial tools, and temporal information, which are processed through a network of collaborative agents. These agents are designed for intelligent query rewriting, context-enriched retrieval from structured threat knowledge bases, and natural language justification of attribution decisions. By combining Retrieval-Augmented Generation (RAG) with Large Language Models (LLMs), AURA enables contextual linking of threat behaviors to known APT groups and supports traceable reasoning across multiple attack phases. Experiments on recent APT campaigns demonstrate AURA’s high attribution consistency, expert-aligned justifications, and scalability. This work establishes AURA as a promising direction for advancing transparent, data-driven, and scalable threat attribution using multi-agent intelligence.
\end{abstract}



\begin{keyword}
Threat Attribution \sep Advanced Persistent Threats (APT) \sep Retrieval-Augmented Generation (RAG) \sep Agentic Systems \sep Large Language Models (LLMs) \sep Cyber Threat Intelligence \sep Tactics Techniques and Procedures (TTPs)


\end{keyword}

\end{frontmatter}

\section{Introduction}
\label{sec:AuraIntroduction}

Attributing cyber threats is a foundational challenge in the field of cybersecurity. Whether for national defense, enterprise protection, or international diplomacy, identifying the actors behind sophisticated attacks is essential for informed response, deterrence, and accountability~\cite{rid2015attributing, egloff2023publicly}. Yet, attribution remains notoriously difficult due to incomplete evidence trails, adversarial deception, and overlapping behavioral signatures across different campaigns~\cite{steffens2020attribution,rani2025comprehensive}. These challenges are especially pronounced in Advanced Persistent Threats (APTs), which are marked by stealth, strategic intent, and persistent targeting of high-value entities. The core difficulty lies not only in determining “who” is responsible, but in doing so accurately, consistently, and transparently based on disparate and often unstructured evidence~\cite{mei2022review,rani2025comprehensive,skopik2020under}.

Cyber threat intelligence reports serve as valuable sources of past attribution signals, often containing rich contextual details such as Tactics, Techniques, and Procedures (TTPs), Indicators of Compromise (IoCs), malware details, adversarial tools, and campaign timelines. However, extracting actionable insight from these artifacts remains largely a manual and error-prone process~\cite{rani2023ttphunter,rani2024ttpxhunter,huang2024mitretrieval,saha2025malaware,cuong2025towards}. Traditional attribution methods, whether based on static heuristics, rule-based indicators, or shallow pattern-matching, fail to capture the nuanced relationships between threat evidence and actor behaviors~\cite{rani2024chasing,rani2025comprehensive,rani2024genesis}. More recent approaches leveraging machine learning and NLP have shown promise, but often lack the ability to reason contextually or to justify their decisions in a way that aligns with expert analysis~\cite{naveen2020deep,noor2019machine,perry2019no}. This limits their trustworthiness, scalability, and operational utility in real-world attribution workflows.

To overcome these challenges, we propose AURA (Attribution Using Retrieval-Augmented Agents), a multi-agent intelligence framework designed to deliver context-aware, interpretable, and knowledge-enhanced threat attribution. AURA ingests a wide range of structured and semi-structured intelligence signals, including TTPs, IoCs, malware artifacts, attacker tools, and campaign timelines, and coordinates a team of specialized agents that collaborate to perform attribution. These agents handle tasks such as query rewriting, context-enriched retrieval, memory management, and justification generation using Large Language Models (LLMs) integrated with Retrieval-Augmented Generation (RAG).

By combining intelligent agent modularity with knowledge-grounded reasoning, AURA bridges the gap between raw threat intelligence and high-level attribution decisions. Unlike prior methods that rely on handcrafted rules or black-box classifiers, AURA generates interpretable outputs by tracing attribution decisions back to contextual evidence within the threat corpus. It enables scalable analysis across campaigns and supports transparency by producing natural language justifications for each attribution decision. This design not only improves attribution accuracy, but also fosters analyst trust and decision support.

\begin{figure}[!h]
    \centering
    \includegraphics[width=12cm, height=3.5cm]{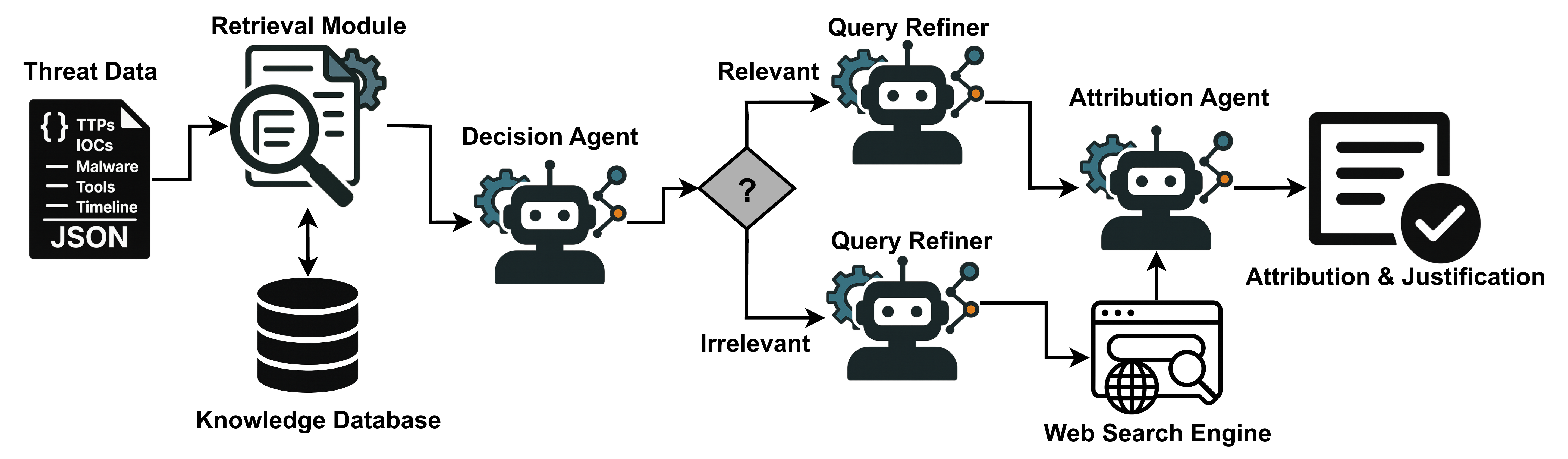} 
    \caption{Overview of AURA: The multi-agent framework comprises specialized agents for query rewriting, knowledge retrieval, and attribution. 
    }
    \label{fig:aura-architecture}
\end{figure}

AURA operates by transforming an analyst's input query into an attribution decision and supporting explanation through a coordinated pipeline of intelligent agents. Conceptually, this process can be framed as a transformation function:
\[
\text{AURA}(Q) = (A, J)
\]
where $Q$ is the natural language query, $A$ is the predicted threat actor, and $J$ is a natural language justification. Each step in this mapping, such as query rewriting, contextual retrieval, actor inference, and explanation, is handled by a specialized agent. An overview of this multi-agent architecture is depicted in Figure~\ref{fig:aura-architecture}. 

We evaluate AURA on a diverse set of real-world APT campaign datasets and demonstrate its effectiveness in producing accurate, human-readable, and context-rich attribution outputs. Our findings indicate that agentic modularity, combined with RAG, significantly improves both the quality and the explainability of cyber threat attribution. In summary, this paper makes the following contributions:




\begin{itemize}
\item We introduce AURA  (Attribution Using Retrieval-Augmented Agents), a multi-agent intelligence framework that enables knowledge-enhanced attribution of cyber threats by integrating Retrieval-Augmented Generation (RAG) with LLMs.
\item We design modular agents for query rewriting, context-aware retrieval, and natural language justification, enabling structured and explainable attribution workflows that produce human-readable, evidence-supported explanations aligned with expert reasoning practices.
\item We perform extensive evaluations of AURA on real-world threat reports, demonstrating its accuracy, robustness, and interpretability across diverse APT scenarios.
\item We develop a chatbot system\footnote{To be made publicly available upon acceptance} based on the proposed AURA framework, aimed at real-world use.
\end{itemize}

The remainder of this paper is structured as follows: Section~\ref{sec:AurarelatedWork} reviews related work. Section~\ref{sec:Aura} introduces the architecture and components of AURA. Section~\ref{sec:AuraExperimentSetup} describes the experimental setup. The results are presented in Section~\ref{sec:AuraResults}, followed by a detailed discussion in Section~\ref{sec:AuraDiscussion}. 
The limitations and future work are discussed in Section~\ref{sec:AuraLimitationAndFutureWork}. Finally, Section~\ref{sec:AuraConclusion} concludes the paper.

\section{Related Work}
\label{sec:AurarelatedWork}



Cyber threat attribution relies on a variety of artifacts, including malware samples, indicators of compromise such as file hashes, IP addresses, and domain names, as well as unstructured threat intelligence reports and behavioral patterns~\cite{rani2025comprehensive}. Earlier approaches concentrated on static features, which often fail to provide reliable attribution when adversaries reuse tools, disguise their activities, or intentionally mislead defenders~\cite{gray2024identifying,rani2025comprehensive}. In response, recent studies have turned toward behavioral characteristics, particularly tactics, techniques, and procedures, as these offer more persistent and meaningful signals for identifying threat actors~\cite{rani2024chasing,noor2019machine}. Researchers have also applied natural language processing and machine learning methods to extract such behavioral indicators from textual threat intelligence, enabling more automated and scalable attribution.

In the context of attributing behavioral patterns, Noor et al.~\cite{noor2019machine} profile threat actors based on the presence of NLP-derived patterns and apply machine learning classifiers. Although effective within a constrained domain, their system lacked generalizability across diverse attack contexts and was trained on a limited set of data samples.  
Irshad and Siddiqui~\cite{irshad2024context} developed an automated pipeline that extracts features such as attack techniques, malware families, and targeted sectors from cyber threat intelligence documents. Their approach employ machine learning classifier on domain-specific embeddings to improve the accuracy and relevance of the extracted information. Their system achieved promising accuracy in classifying threat actors, but it lacked contextual reasoning or explainability. Building on the growing focus on behavioral attribution, Rani et al.~\cite{rani2024chasing} proposed a structured method that organizes MITRE ATT\&CK tactics, techniques, and procedures into kill chain phases and compares these sequences against known actor profiles using a novel similarity measure. While this approach enables pattern-based attribution across campaigns, it assumes reliable TTP extraction and lacks support for reasoning over incomplete or mixed evidence sources. To model behavioral patterns, Böge et al. \cite{boge2025unveiling} proposed a hybrid architecture combining transformers and convolutional networks to analyze sequences of commands executed by threat actors. Their introduction of a standardized command language improved robustness across varied data distributions. However, the approach operated exclusively on command logs and did not incorporate structured threat knowledge or support broader contextual analysis.

For malware-based attribution, Rosenberg et al.~\cite{rosenberg2017deepapt} introduce DeepAPT, a deep learning approach that uses raw dynamic malware behavior for APT attribution. Malware samples are executed in a sandbox to generate behavior reports, and the words in these reports are treated as features. These are processed as natural language inputs and encoded to train a deep neural network for classification. Rani et al.~\cite{rani2024genesis} focused on malware-based APT attribution by extracting static, dynamic, and temporal features from malware samples and training machine learning classifiers to identify APT groups. While these approaches effectively leverage malware artifacts, it do not incorporate structured reasoning or contextual intelligence beyond malware behaviors. 


To leverage threat reports for attribution, the NO-DOUBT system~\cite{perry2019no} employs a weakly supervised BERT-based classifier trained on cyber threat intelligence reports to generate attribution scores. Although scalable, the system lacked interpretability and did not support semantic retrieval or reasoning. Guru et al. \cite{guru2025technique} proposed an end-to-end pipeline combining GPT-4 for extracting techniques and OpenAI embeddings for matching with known actor profiles. However, their framework operated in a single-pass fashion, treating LLMs as extractors rather than reasoning agents, and lacked modular design or justification synthesis. Naveen et al.~\cite{naveen2020deep} propose a deep learning framework that attributes threat actors from unstructured CTI reports using domain-specific neural embeddings. Their SIMVER representation encodes semantic similarity between words, enabling a dense neural network to better than traditional methods in attributing APT groups based on textual TTP patterns.


While existing approaches have made significant progress by utilizing individual artifacts such as malware behavior, threat reports, or sequences of tactics, techniques, and procedures, they often treat these sources in isolation without integrating them into a unified analysis. In addition, many models operate as opaque systems with limited transparency in how attribution decisions are made. The lack of clear reasoning and justification further reduces trust and interpretability for human analysts. These limitations highlight the need for attribution methods that combine multiple types of evidence while offering explainable outputs supported by structured reasoning.

To bridge these gaps, we introduce AURA (Attribution Using Retrieval Augmented Agents), a modular and explainable framework designed for real-world cyber threat attribution. AURA combines structured threat data, semantic retrieval, and reasoning powered by large language models to unify diverse intelligence artifacts into a coherent attribution process. The framework is composed of specialized agents, each responsible for a distinct stage in the pipeline, including input processing, query rewriting, semantic search, attribution generation, and justification synthesis. This agent-based architecture supports flexible coordination and scalability across varied input formats.

AURA generates attribution results along with natural language justifications, enhancing both transparency and analyst trust. Its capability to perform dynamic reasoning over multiple types of threat intelligence, including tactics, techniques and procedures, malware behavior, and unstructured reports, makes it a robust solution for complex attribution scenarios. A comparative overview of related approaches is provided in Table~\ref{tab:comparison}, which highlights AURA's distinctive combination of structured data extraction, semantic alignment, and modular reasoning over diverse inputs.

\begin{sidewaystable}[htbp]
\centering
\caption{Comparison of AURA with Existing Threat Attribution Methods}
\label{tab:comparison}
\begin{tabular}{|p{2.2cm}|p{2.6cm}|p{3.2cm}|p{3cm}|p{2cm}|p{3.2cm}|p{2.3cm}|}
\hline
\textbf{Method} & \textbf{TTP Extraction} & \textbf{Attribution Method} & \textbf{Explainability} & \textbf{LLM Use} & \textbf{Heterogeneous Input} & \textbf{Modular Design} \\
\hline
\hline
Noor et.al~\cite{noor2019machine} &  \cmark (from threat reports) & Deep Neural network & \xmark & \xmark & \xmark & \xmark \\
\hline
Rani et. al~\cite{rani2024chasing} &\cmark (from threat reports) & Graph similarity over campaigns & \xmark & \xmark & \xmark & \xmark \\
\hline
Rosenberg et. al~\cite{rosenberg2017deepapt}  & \xmark & Deep Neural Network & \xmark & \xmark & \xmark & \xmark \\
\hline
Irshad et. al~\cite{irshad2024context} &\cmark(from malware reports) & Rule-based matching to MITRE &\cmark(basic feature names using Using LIME~\cite{ribeiro2016should}) & \xmark & \xmark & \xmark \\
\hline
Rani et. al~\cite{rani2024genesis} &\cmark(from malware+TTPs) & Supervised ML classifiers & \xmark & \xmark & \xmark & \xmark \\
\hline
NO-DOUBT~\cite{perry2019no} &\xmark & Weakly supervised BERT model & \xmark & \xmark & \xmark & \xmark \\
\hline
Guru et al.~\cite{guru2025technique} &\cmark(via GPT + Embeddings) & Vector similarity with actor profiles & \xmark &\cmark(GPT) & Partial (no structured heterogeneity) & \xmark \\
\hline
Böge et al.~\cite{boge2025unveiling} & \xmark (uses command logs) & CNN + Transformer hybrid & \xmark & \xmark & \xmark & \xmark \\
\hline
Naveen et al.~\cite{naveen2020deep}  & \xmark & Deep Neural Network & \xmark & \xmark & \xmark & \xmark \\
\hline
\textbf{AURA (Ours)} &\cmark\cmark(from threat reports) & LLM-based reasoning and generation & \cmark\cmark (natural language justification) & \cmark\cmark (multi-agent LLM) & \cmark\cmark (heterogeneous inputs with retrieval) & \cmark\cmark (agent-based architecture) \\
\hline
\end{tabular}
\end{sidewaystable}

\section{AURA}
\label{sec:Aura}

In this section, we describe the design and components of AURA (Attribution Using Retrieval-Augmented Agents), a multi-agent intelligence framework for performing knowledge-enhanced, interpretable, and scalable cyber threat attribution. AURA orchestrates specialized agents to process diverse cyber threat signals, including TTPs, IoCs, malware details, attacker tools, and campaign timelines, and performs attribution by integrating structured retrieval with the reasoning capabilities of Large Language Models (LLMs). The framework leverages Retrieval-Augmented Generation (RAG) within an agentic architecture to facilitate dynamic query transformation, contextual retrieval, actor inference, and natural language justification. This modular, agent-based design is inspired by recent LLM-driven systems such as MalGEN~\cite{saha2025malgengenerativeagentframework,guo2024large,wu2023autogen}, which employ coordinated agent workflows to generate and reason about behaviorally diverse malware, highlighting the broader applicability of agentic reasoning in cybersecurity.

\subsection{Overview of the Framework}

\begin{figure}[h]
  \centering
  \includegraphics[width=9cm, height=10cm]{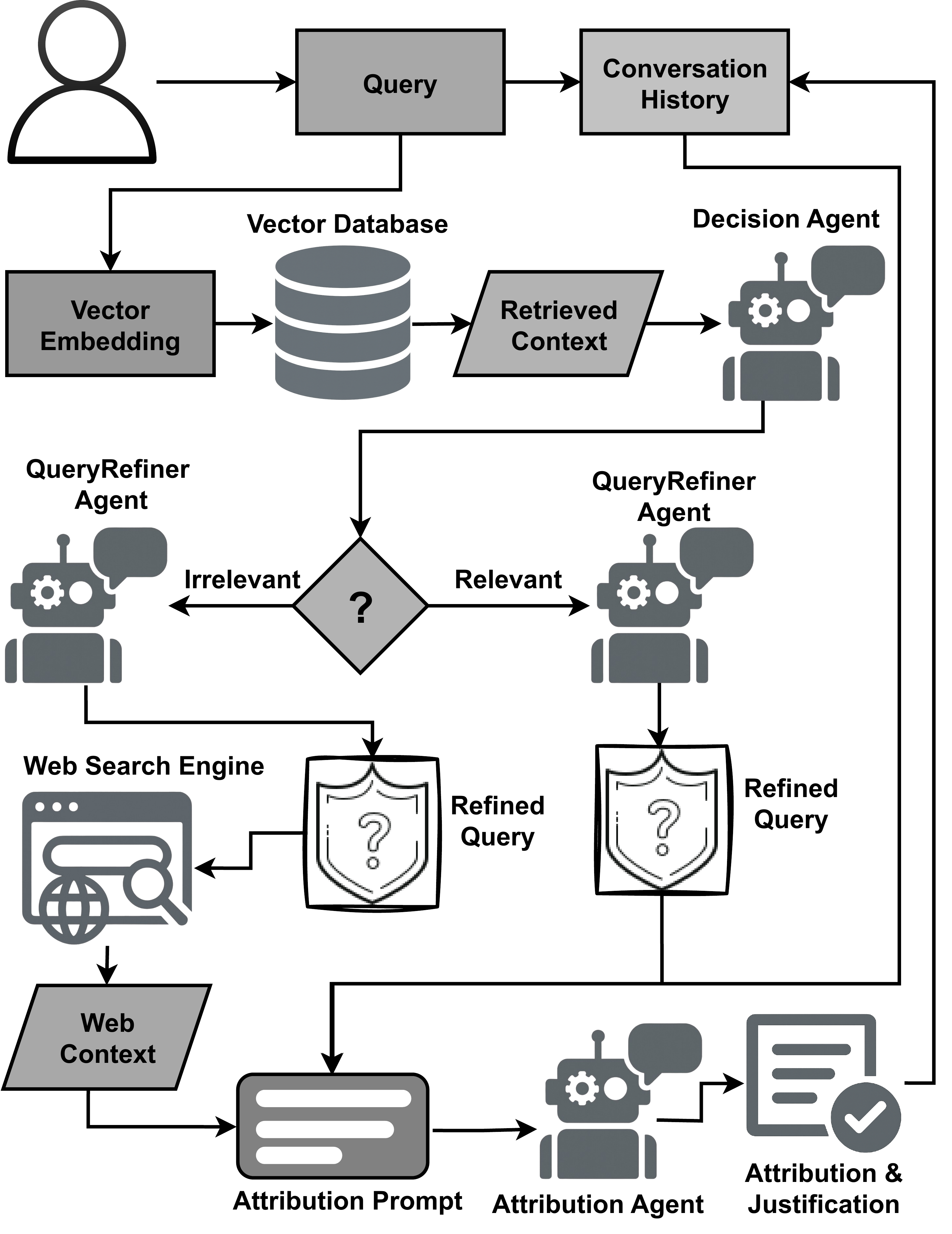}
  \caption{Architecture of the AURA-based chat-bot system.}
  \label{fig:aura_architecture}
\end{figure}

AURA is architected as a modular, multi-agent system that processes analyst prompts or threat intelligence queries in a coordinated pipeline composed of six key components: (i) Input and Preprocessing (ii) Semantic Retriever (iii) Decision Agent (iv) Query Rewriting Agent (v) Web Search Engine Module (vi) Attribution Generation Agent (vii) Conversational Memory Module. These agents interact via structured prompts and shared memory, enabling rich context accumulation and knowledge-enhanced attribution. The high-level architecture of AURA-based chat-bot system is illustrated in Figure~\ref{fig:aura_architecture}. The notations and their corresponding descriptions used throughout the methodology are summarized in Table~\ref{tab:notation}.

\begin{table}[h!]
\centering
\begin{tabular}{ll}
\hline
\textbf{Notation} & \textbf{Description} \\
\hline
$Q$ & Input analyst query \\
$Q'$ & Rewritten, semantically precise query \\
$\mathcal{L}$ & Space of natural language queries \\
$\mathcal{E}$ & Extracted threat entities (TTPs, IoCs, etc.) \\
$\mathcal{M}$ & Conversational memory (historical context) \\
$\mathcal{C}$ & Threat intelligence corpus \\
$C$ & Top-$k$ retrieved chunks from corpus \\
$\mathcal{A}$ & Set of known threat actors \\
$A$ & Predicted threat actor \\
$J$ & Natural language justification \\
$f_{\text{pre}}$ & Preprocessing function \\
$f_{\text{rew}}$ & Query rewriting function \\
$f_{\text{ret}}$ & Semantic retrieval function \\
$f_{\text{attr}}$ & Attribution generation function \\
$f_{\text{just}}$ & Justification synthesis function \\
$f_{\text{mem}}$ & Memory update function \\
$f_{\text{embed}}$ & Embedding generation function \\
\hline
\end{tabular}
\caption{Notation used in the AURA pipeline}
\label{tab:notation}
\end{table}

\subsection{Input and Preprocessing}


AURA accepts as input either natural language queries from analysts or parsed threat intelligence content (e.g., extracted from reports).These inputs often reference a mixture of data, such as \textit{TTPs, IoCs, malware names, tool usage, or temporal patterns}. A lightweight preprocessing module extracts metadata using state-of-the-art LLM, identifying TTPs, actor names, infrastructure, time ranges, and malware/tool mentioned. Let the input query be denoted by $Q \in \mathcal{L}$, where $\mathcal{L}$ is the space of natural language queries. The structured threat entities extracted from $Q$ are denoted by $$\mathcal{E} = f_{\text{pre}}(Q)$$ where $\mathcal{E}$ includes elements like TTPs, IoCs, malware details, and timeline stored in a json file.

These inputs are stored in a persistent conversational memory $\mathcal{M}$ for use across the attribution workflow and to support inter-query contextualization.

\subsection{Semantic Retriever}


The refined query is passed to a \textit{Semantic Retriever Agent} that performs vector-based retrieval over a \textit{knowledge-enhanced corpus} of structured and semi-structured cyber threat intelligence reports. This corpus is denoted as $\mathcal{C}$, and the top-$k$ retrieved chunks as $$C = \{c_1, c_2, \dots, c_k\} = f_{\text{ret}}(Q) \subset \mathcal{C}$$ AURA uses the vector database, indexing dense embeddings generated via OpenAI text embedding model. Each chunk is ranked using cosine similarity:
\[
\text{sim}(Q, c_i) = \frac{f_{\text{embed}}(Q) \cdot f_{\text{embed}}(c_i)}{\|f_{\text{embed}}(Q)\| \cdot \|f_{\text{embed}}(c_i)\|}
\]
ensuring that attribution reasoning is grounded in relevant, contextual threat knowledge.

\subsection{Decision Agent}
Since the context is retrieved using vector similarity, its relevance to the objective may vary. Passing irrelevant context to the attribution agent can distract or mislead the LLM, potentially resulting in misattribution. To address this, we integrate a decision agent that evaluates the retrieved context before it is passed to the attribution agent. This agent is prompted to determine whether the context is relevant to the final attribution objective.

\subsection{Query Rewriting Agent}


To handle ambiguities and ensure semantic precision, AURA employs a \textit{Query Rewriting Agent}. This agent refines the analyst's prompt using the conversational history and previously extracted entities. For example, vague references such as ``they" or ``this group" are resolved to explicit actor names like \texttt{APT28} or \texttt{Lazarus Group}. The rewritten query $Q'$ is obtained as $$Q' = f_{\text{rew}}(Q, \mathcal{E}, \mathcal{M})$$ This ensures that the reformulated query is unambiguous, aligned with the objective task, and optimized for semantic retrieval, particularly in cases involving follow-up interactions or coreference resolution.

\subsection{Web Search Engine Module}

If the decision agent determines that the initially retrieved context is irrelevant, an external web search is initiated to gather more suitable information from publicly available sources. The query is first reformulated in a tailored manner to enhance the relevance of the search results. The new information obtained through this process is then provided to the attribution agent to support the final decision-making.

\subsection{Attribution Generation Agent}

Using the retrieved evidence and the rewritten query, the \textit{Attribution Generation Agent} invokes a LLM along with retrieved context to identify the most probable threat actor and generate a natural language justification for its decision. It aligns observed TTPs, malware/tool usage patterns, and temporal indicators with known actor profiles to compute the predicted actor $A$ as:

\begin{equation}
A = f_{\text{attr}}(Q', \mathcal{E}, C)
\end{equation}

where $A \in \mathcal{A}$ and $\mathcal{A}$ denotes the set of known threat actors. Simultaneously, it produces a justification $J$ for this decision as:

\begin{equation}
J = f_{\text{just}}(A, \mathcal{E}, C)
\end{equation}

where $J \in \mathcal{L}$. The generated justification synthesizes retrieved evidence, highlights aligned TTPs and temporal patterns, and compares them to historical behaviors of the predicted actor. This step leverages LLM reasoning over retrieval-augmented input to deliver interpretable, evidence-backed attributions, even in scenarios involving overlapping or ambiguous indicators.




\subsection{Conversational Memory Module}

AURA maintains a \textit{conversational memory}, integrated into its chatbot system, to track prior queries, attribution decisions, and justifications across multiple turns. This enables coherent multi-turn interactions by preserving contextual continuity and ensuring that follow-up queries are interpreted in light of previous inputs and outputs. The memory buffer is updated as:

\begin{equation}
\mathcal{M}' = f_{\text{mem}}(\mathcal{M}, Q', \mathcal{E}, A, J)
\end{equation}

where $\mathcal{M}$ represents the current memory state, and the updated state $\mathcal{M}'$ integrates the rewritten query $Q'$, retrieved evidence $\mathcal{E}$, predicted actor $A$, and generated justification $J$. This facilitates consistent dialogue grounding and enhances the chatbot’s ability to support evolving analyst queries within a session.

\paragraph{Final Output}  
The complete output of the AURA pipeline is the attribution decision $A$ and its justification $J$, represented as:
\[
\text{AURA}(Q) = (A, J)
\]

This formulation illustrates how AURA decomposes attribution into modular reasoning steps while maintaining traceability through extracted knowledge entities $\mathcal{E}$.

\subsection{Implementation Details}

AURA is implemented in Python, using the LangChain framework for agent orchestration and memory management, Qdrant for vector database for storing purpose and similarity search, and LLMs from OpenAI and Anthropic for generation and reasoning. The LLM for all agents, except the final attribution agent, is GPT-4o. We replace the final attribution agent for each model-specific experiment. The system is modular, supporting plug-and-play replacement of individual agents or embedding models. Communication between agents is handled via structured function-calling protocols, making the framework \textit{extensible, provider-agnostic, and scalable} for deployment across different cyber intelligence environments.

\section{Experiments Setup}
\label{sec:AuraExperimentSetup}


\subsection{Dataset}
\label{subsec:AuraDataset}
\begin{figure}
    \centering
    \includegraphics[width=\linewidth]{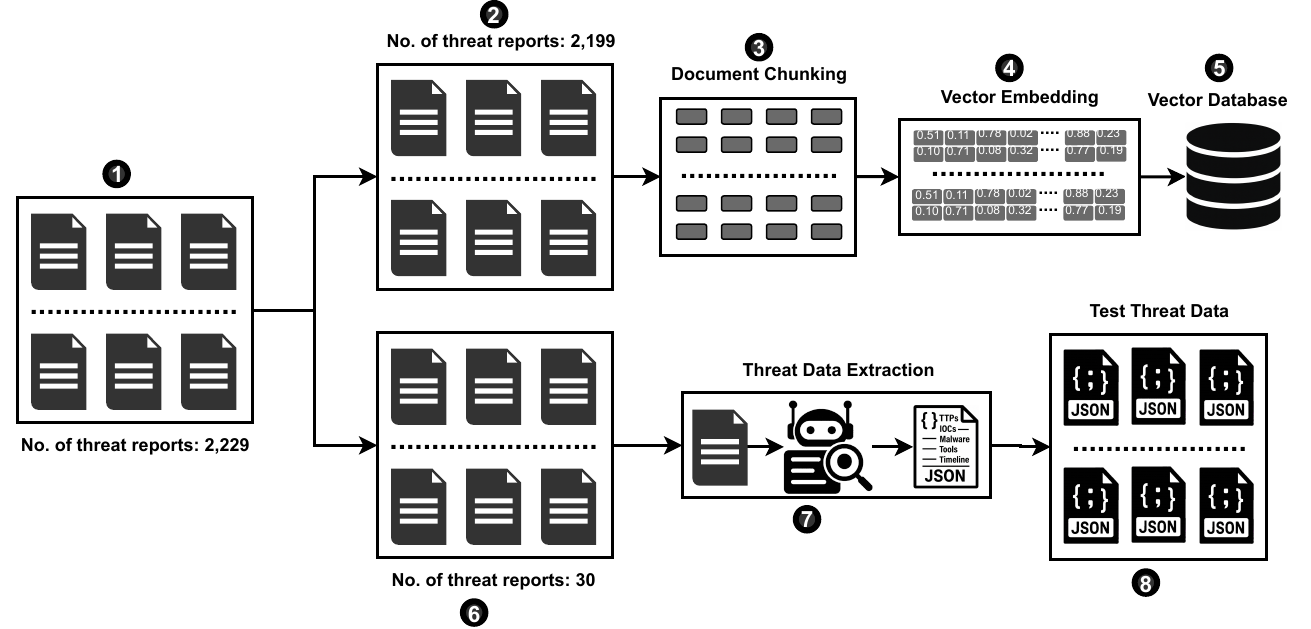}
    \caption{Dataset Preparation}
    \label{fig:AuraDataset}
\end{figure}

Effective retrieval augmentation requires a substantial repository of task-specific knowledge to support the agents during attribution. To build this knowledge base, we collect threat analysis reports published by reputable cybersecurity firms such as Google, CrowdStrike, Kaspersky, and others. The dataset is sourced from publicly available repositories on GitHub~\cite{aptnotes,aptcybercriminal}, comprising a total of 2,229 threat reports (Step \tikz[baseline=(char.base)]{
  \node[shape=circle, fill=black, text=white, inner sep=1pt, minimum size=1em] (char) {\textbf{1}};
} in Fig~\ref{fig:AuraDataset}).

To mitigate any bias from model pretraining data, we split the dataset based on the knowledge cutoff dates of the LLMs. Specifically, 2,199 reports are used to populate the vector database that serves as AURA's knowledge base (Step \tikz[baseline=(char.base)]{
  \node[shape=circle, fill=black, text=white, inner sep=1pt, minimum size=1em] (char) {\textbf{2}};
} in Fig~\ref{fig:AuraDataset}), while the remaining 30 reports are reserved as a held-out test set (Step \tikz[baseline=(char.base)]{
  \node[shape=circle, fill=black, text=white, inner sep=1pt, minimum size=1em] (char) {\textbf{6}};
} in Fig~\ref{fig:AuraDataset}). These test reports are used to extract attack-related artifacts, which are then passed into the AURA framework as input for threat actor attribution and justification generation (Step \tikz[baseline=(char.base)]{
  \node[shape=circle, fill=black, text=white, inner sep=1pt, minimum size=1em] (char) {\textbf{7}};
} in Fig~\ref{fig:AuraDataset}). The overall process for curating the knowledge base and generating structured test data is illustrated at Step \tikz[baseline=(char.base)]{
  \node[shape=circle, fill=black, text=white, inner sep=1pt, minimum size=1em] (char) {\textbf{8}};
} in Fig~\ref{fig:AuraDataset}.

As threat reports are often multi-page documents and LLMs have limitations on context length, we divide each report in the knowledge base into smaller, manageable chunks (Step \tikz[baseline=(char.base)]{
  \node[shape=circle, fill=black, text=white, inner sep=1pt, minimum size=1em] (char) {\textbf{3}};
} in Fig~\ref{fig:AuraDataset}). To preserve the semantic flow across chunks, we maintain an overlap of $50$ tokens between consecutive segments. After chunking, we compute vector embeddings (Step \tikz[baseline=(char.base)]{
  \node[shape=circle, fill=black, text=white, inner sep=1pt, minimum size=1em] (char) {\textbf{4}};
} in Fig~\ref{fig:AuraDataset}) for each chunk and store them in a vector database (Step \tikz[baseline=(char.base)]{
  \node[shape=circle, fill=black, text=white, inner sep=1pt, minimum size=1em] (char) {\textbf{5}};
} in Fig~\ref{fig:AuraDataset}). These stored embeddings serve as the knowledge base for the attribution framework, allowing relevant information to be retrieved based on the similarity between the query's embedding and the stored vectors.


Since threat reports are originally in unstructured textual format, they are well-suited for use as a natural language knowledge base. However, in real-world scenarios, analysts may not always have access to detailed textual reports. Instead, threat data may be available in structured formats such as JSON or CSV. To simulate this practical setting, we use the \texttt{gpt-4o} model to extract structured threat indicators such as TTPs, IoCs, malware details, tools, and attack timelines from the textual test reports into a structured JSON format. This conversion ensures that the test input mimics realistic machine-readable threat data, while preserving key information required for attribution. 



\subsection{LLM Model Selection}
\label{subsec:AuraLLMModelSelection}
For this analysis, we focus on black-box LLMs because of their state-of-the-art reasoning abilities, effectiveness in correlating complex evidence, and ability to generate well-structured responses. We evaluate four proprietary models from OpenAI and Anthropic: \texttt{gpt-4o}, \texttt{gpt-4o-mini}, \texttt{Claude 3.5 Haiku}, and \texttt{Claude 3.5 Sonnet}.




\subsection{Experiment}
\label{subsec:AuraExperiment}

Since AURA is capable of performing real-time web-based retrieval, there is a possibility that threat data from the test set, even though it is historical, might be available somewhere on the internet. To ensure a fair and controlled evaluation, and to avoid any potential data leakage, we disable the web search capability of the underlying LLMs during testing. The results discussed in Section~\ref{sec:AuraResults} reflect the performance of AURA when operating solely on its internal knowledge base, without accessing any external search engine. This setup provides a conservative baseline; we anticipate that performance would further improve when tested on proprietary or previously unseen threat intelligence data, where retrieval-augmented LLMs can fully leverage external sources.

In addition, we align our experimental setup with the \textit{4C attribution framework}~\cite{steffens2020attribution}, which defines attribution granularity levels. According to this model, the highest level of attribution granularity involves identifying specific individuals or organizations, while the second-highest involves attributing an attack to a nation-state. To support both levels of granularity, our framework is configured to attribute threats at the level of known threat groups as well as the possible linked nations.

Incorporating nation-level attribution enhances the framework’s utility, especially in cases where threat actors operate collaboratively under a common national interest or share similar modus operandi. This dual-granularity approach allows AURA to identify the most likely responsible group and also infer geopolitical context, thereby improving the depth and relevance of the attribution analysis.

Due to adversarial deception and overlapping modus operandi, attribution is not always definitive~\cite{steffens2020attribution,rani2024genesis,rani2024chasing}. To assess AURA's robustness under such uncertainty, we extend the evaluation beyond top-1 attribution (most likely threat group) to include top-2 attribution, capturing the two most plausible actors. This accounts for cases where multiple threat groups exhibit similar behavioral patterns. Additionally, to accommodate the variability in LLM outputs, we evaluate AURA using the widely adopted \textit{pass@3} metric~\cite{chen2021evaluating,kulal2019spoc}, which measures whether a correct attribution appears in any of the top three generations.

\section{Results}
\label{sec:AuraResults}

\begin{figure}[h]
    \centering

    \begin{subfigure}[b]{0.48\textwidth}
        \includegraphics[width=\textwidth]{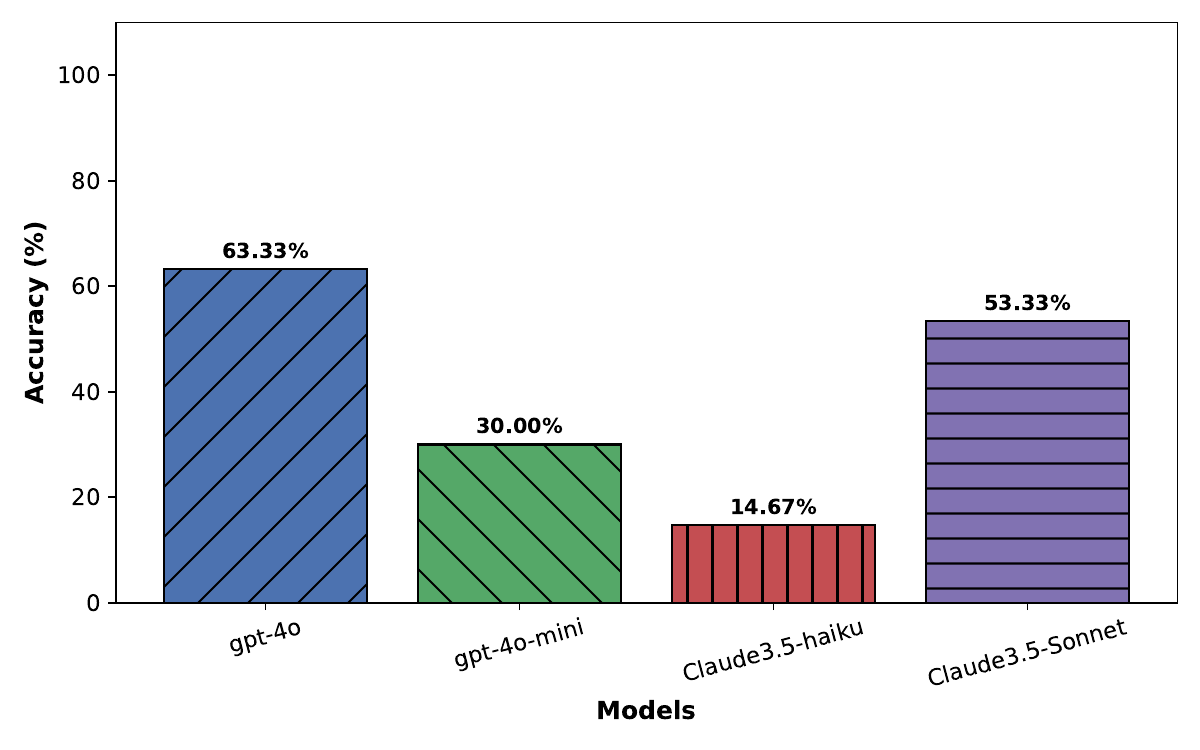}
        \caption{Top-1 Group-wise Performance}
    \end{subfigure}
    \hfill
    \begin{subfigure}[b]{0.48\textwidth}
        \includegraphics[width=\textwidth]{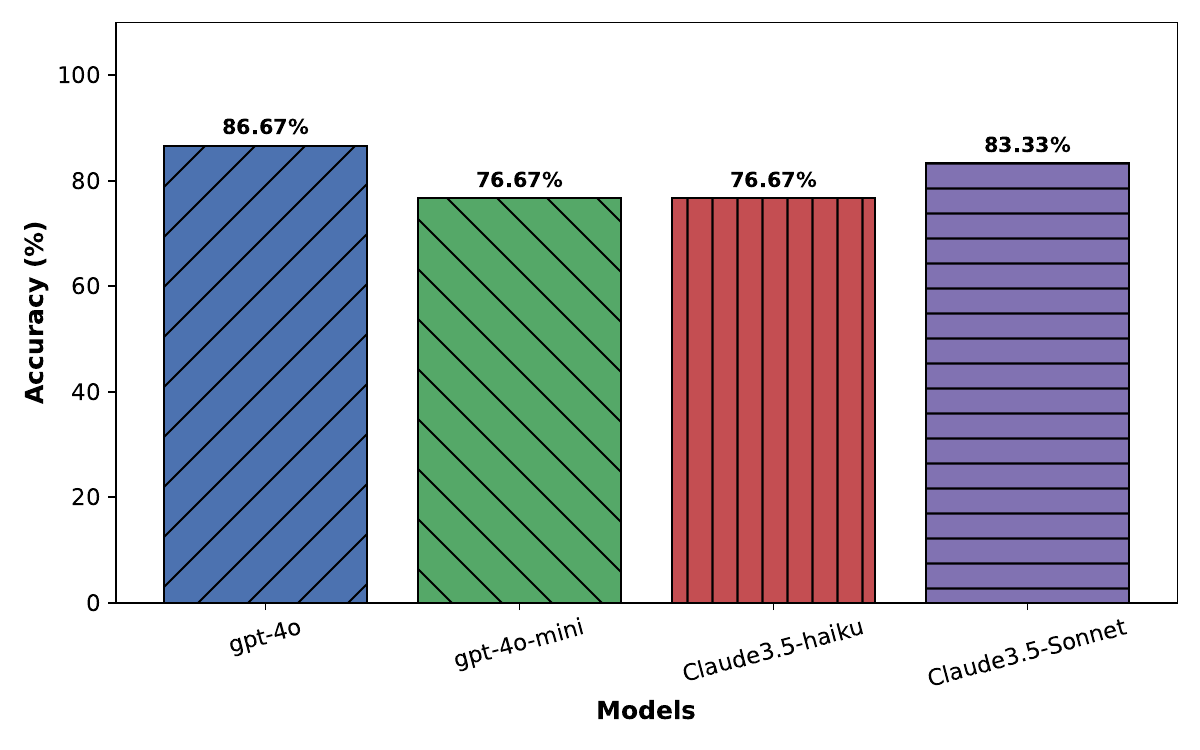}
        \caption{Top-1 Nation-wise Performance}
    \end{subfigure}

    \begin{subfigure}[b]{0.48\textwidth}
        \includegraphics[width=\textwidth]{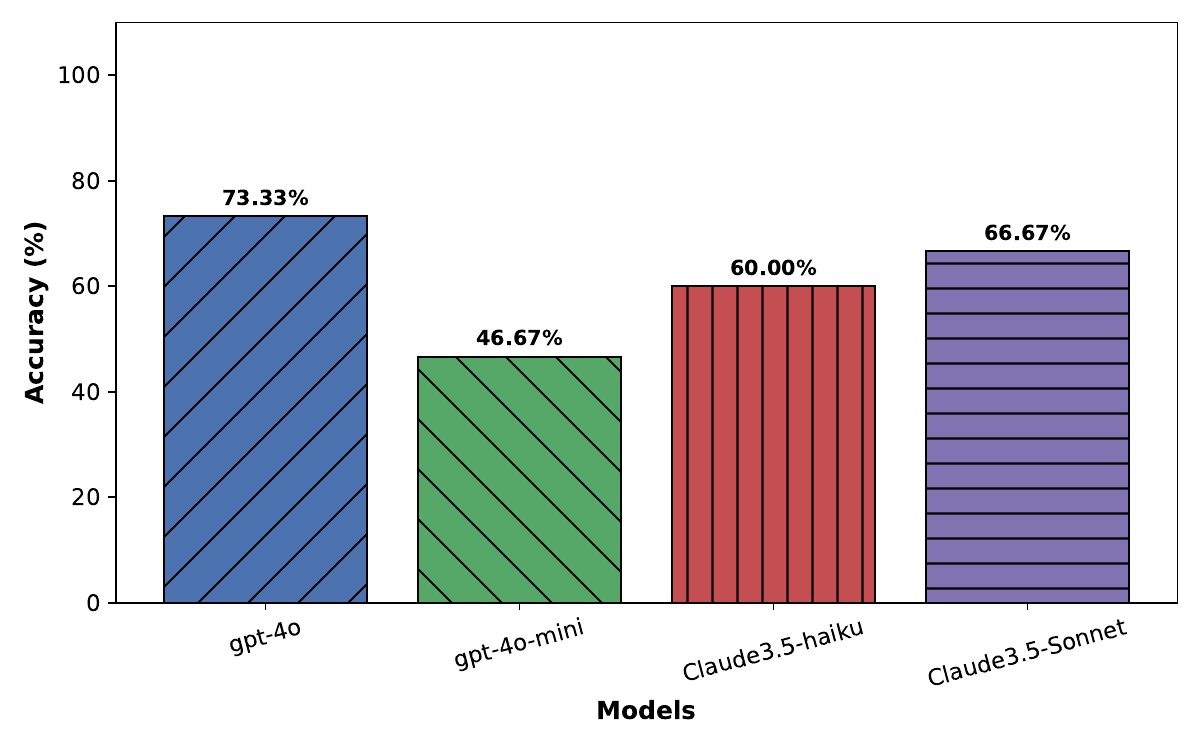}
        \caption{Top-2 Group-wise Performance}
    \end{subfigure}
    \hfill
    \begin{subfigure}[b]{0.48\textwidth}
        \includegraphics[width=\textwidth]{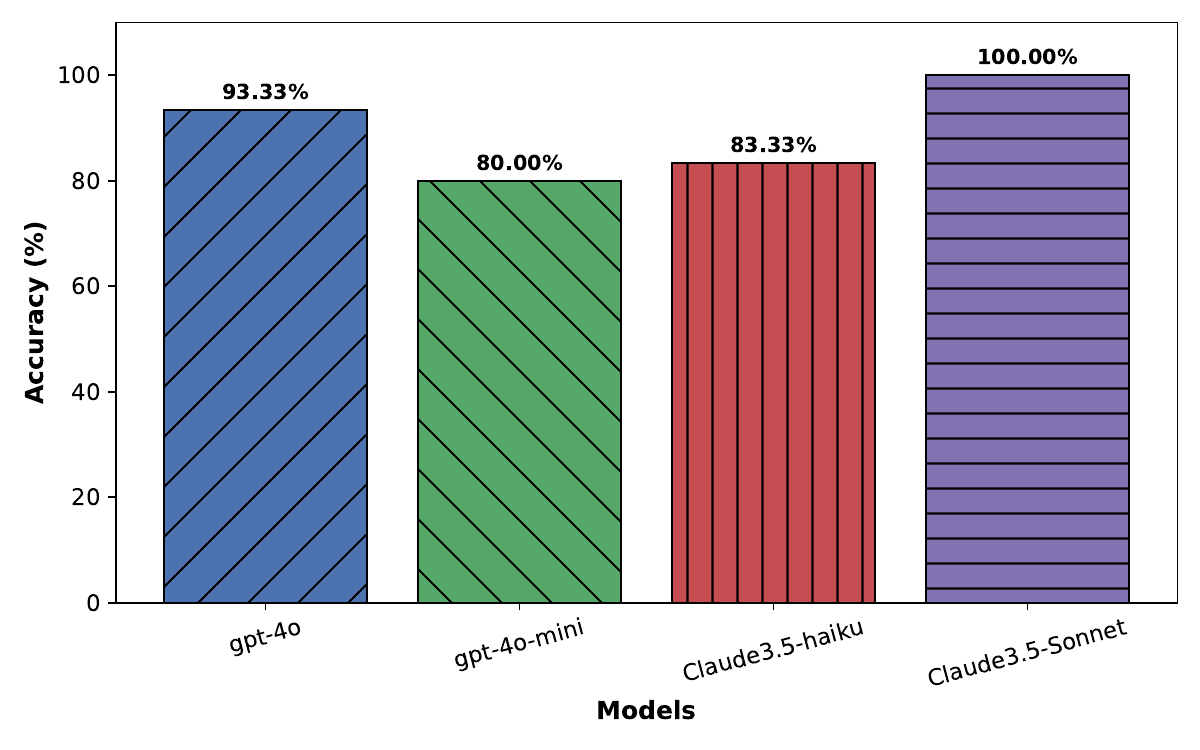}
        \caption{Top-2 Nation-wise Performance}
    \end{subfigure}

    \caption{Attribution accuracy of four LLMs across group-wise and nation-wise levels under top-1 and top-2 settings.}
    \label{fig:individual_attribution_plots}
\end{figure}

This section presents the performance of four black-box LLMs—\texttt{gpt-4o}, \texttt{gpt-4o-mini}, \texttt{Claude 3.5 Haiku}, and \texttt{Claude 3.5 Sonnet}—in performing threat attribution across two granular levels: \textit{group-wise} and \textit{nation-wise}. The evaluation is carried out under both top-1 and top-2 ranking settings. Figure~\ref{fig:individual_attribution_plots} provides a comparative visualization of the accuracy across these different settings.


\textbf{Group-wise Attribution.} For group-level attribution, \texttt{gpt-4o} achieves the highest performance, with a top-1 accuracy of 63.33\% and a top-2 accuracy of 73.33\%. This demonstrates the model’s ability to correctly identify the responsible threat group either as the top candidate or among the top two predictions. \texttt{Claude 3.5 Sonnet} also performs competitively, achieving 53.33\% top-1 and 66.67\% top-2 accuracy. Notably, these predictions are made from a large label space comprising over 150+ known threat groups as documented by MITRE ATT\&CK. The ability to narrow down to the correct group from such a wide range of possibilities underscores the effectiveness of the models. Furthermore, all models show a consistent improvement from top-1 to top-2 accuracy, showing the practical value of allowing multiple candidates in scenarios where attribution may be ambiguous.

\textbf{Nation-wise Attribution.} Nation-level attribution yields significantly higher accuracy across all models. \texttt{Claude 3.5 Sonnet} reaches 83.33\% top-1 accuracy and a perfect 100\% under the top-2 setting, demonstrating its strong alignment with geopolitical patterns in threat data. \texttt{gpt-4o} also performs well, achieving 86.67\% and 93.33\% for top-1 and top-2 respectively. The overall performance in nation-level attribution indicates that even when threat group identification is challenging, LLMs are capable of inferring broader national affiliations based on behavioral indicators and contextual evidence.

These results validate the design of the AURA framework, particularly the benefits of retrieval augmentation, query rewriting, and justification synthesis in improving attribution performance. The observed gains from top-1 to top-2 further demonstrate that LLMs are able to surface multiple plausible candidates, which is particularly useful in complex or ambiguous scenarios often encountered in threat intelligence workflows.



    

\section{Discussion}
\label{sec:AuraDiscussion}


In this section, we analyze how the experimental findings validate the effectiveness of the AURA framework for cyber threat attribution. The results presented in Section~\ref{sec:AuraResults} demonstrate that AURA can accurately attribute threats by leveraging structured threat data in conjunction with retrieval-augmented reasoning. The framework achieves up to 63.33\% top-1 and 73.33\% top-2 accuracy at the group level, suggesting that the integration of semantic retrieval with task-specific knowledge effectively grounds the attribution process. Furthermore, the even higher performance in nation-level attribution highlights AURA’s ability to correlate retrieved context with broader attribution patterns observed in real-world campaigns.
The modular design of AURA contributes significantly to attribution quality. The query rewriting agent helps resolve ambiguity, improving retrieval precision. Context-aware retrieval ensures that relevant and specific evidence is surfaced for each query, which the reasoning agent then uses for attribution generation. 
The incorporation of conversational memory and the generation of natural language justifications contributed to making AURA suitable for real-world analyst workflows.

\subsection{Generated Justification Assessment}
We performed a comprehensive evaluation of the natural language justifications generated by AURA’s synthesis agent using two complementary approaches: (i) automated linguistic and semantic metrics, and (ii) a human-aligned evaluation performed by a language model (LLM-as-Judge).

\subsubsection{Automated Evaluation}
\begin{figure}[!h]
    \centering

    \begin{subfigure}[b]{0.48\textwidth}
        \includegraphics[width=\textwidth]{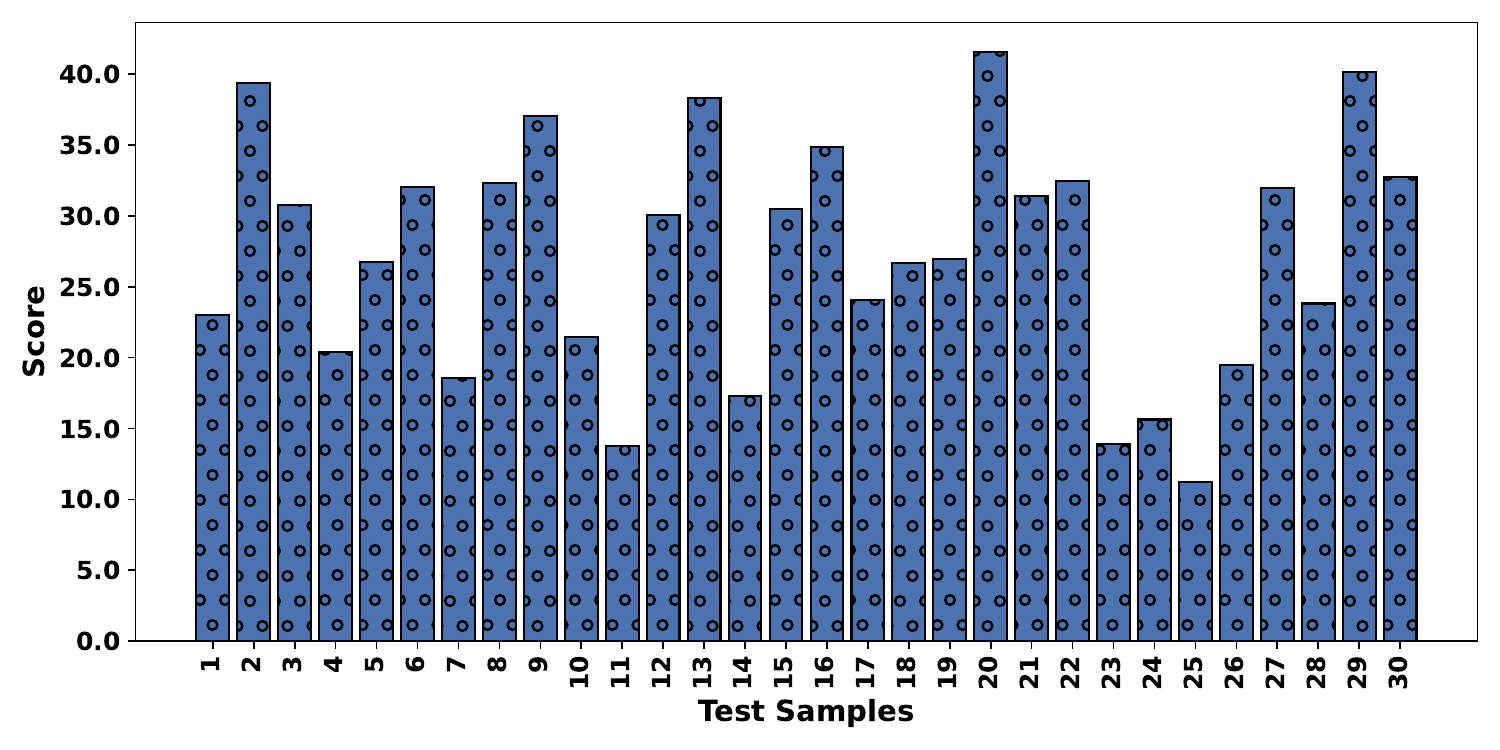}
        \caption{Readability}
    \end{subfigure}
    \hfill
    \begin{subfigure}[b]{0.48\textwidth}
        \includegraphics[width=\textwidth]{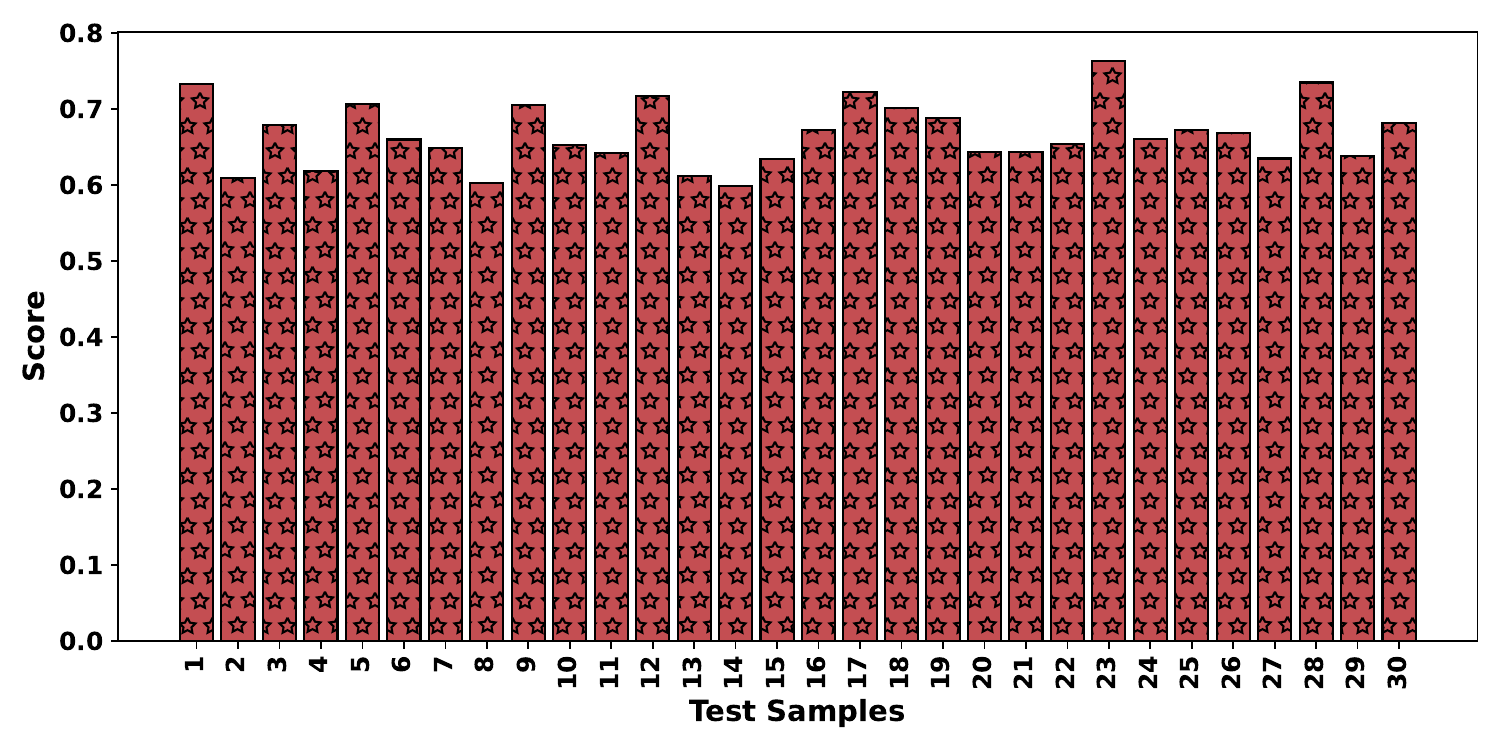}
        \caption{Lexical Richness (TTR)}
    \end{subfigure}

    \begin{subfigure}[b]{0.48\textwidth}
        \includegraphics[width=\textwidth]{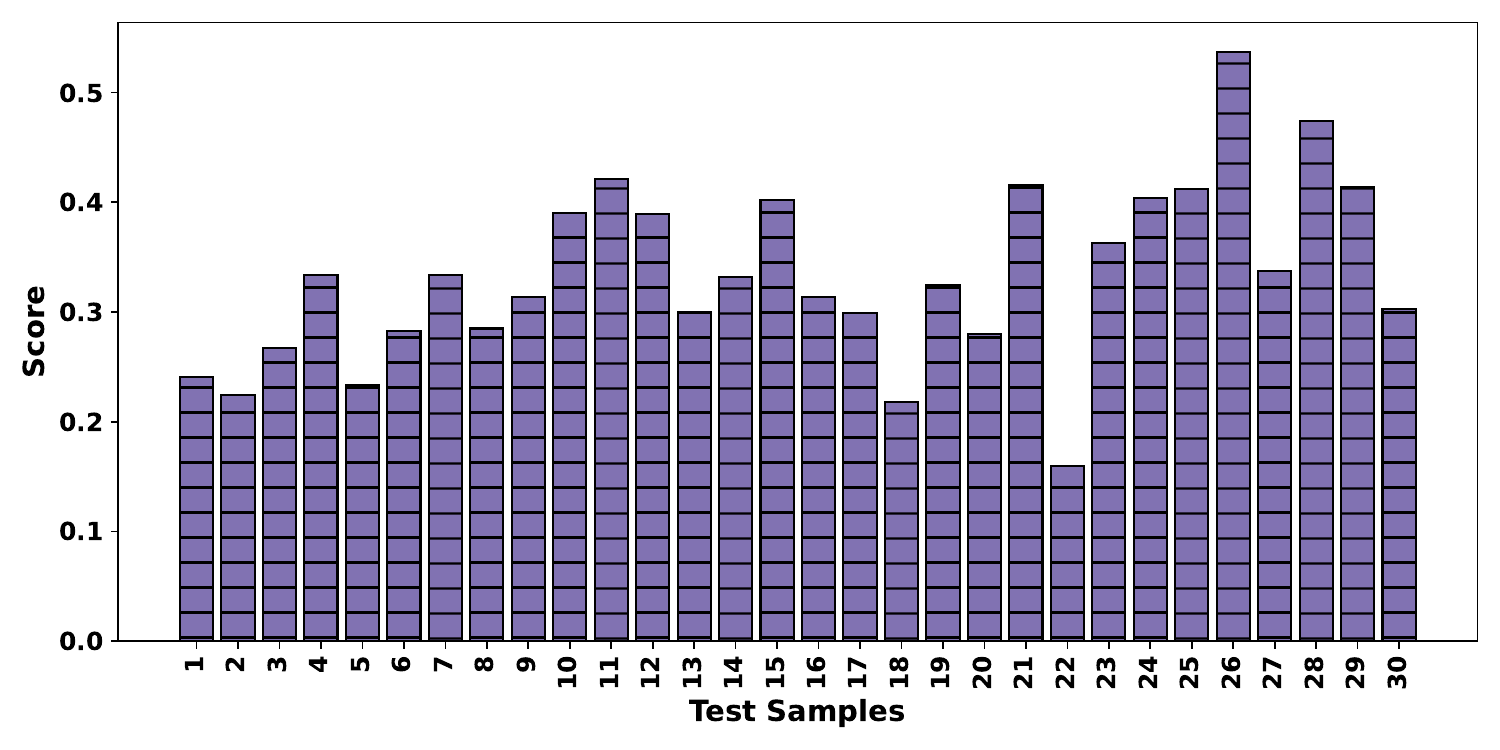}
        \caption{Embedding Coherence}
    \end{subfigure}
    \hfill
    \begin{subfigure}[b]{0.48\textwidth}
        \includegraphics[width=\textwidth]{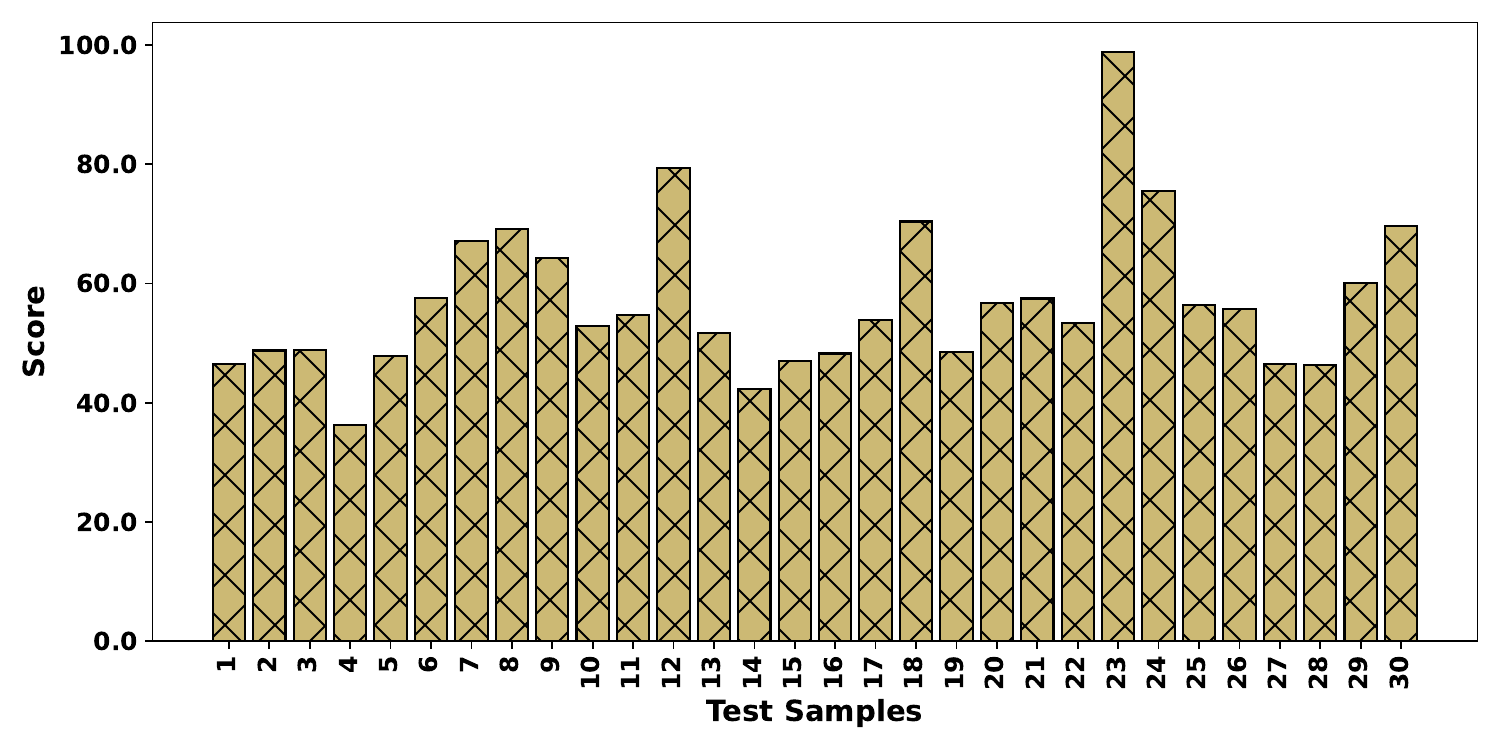}
        \caption{Perplexity Score}
    \end{subfigure}

    \caption{Evaluation of justification quality using four linguistic and semantic measures. Each bar represents the raw score for an individual justification, visualized using unique colors and hatch patterns for clarity.}
    \label{fig:justification_metrics_subfigs}
\end{figure}

We assessed justifications using four widely adopted measures: readability (Flesch Reading Ease), lexical richness (Type-Token Ratio), semantic coherence (sentence-level embedding similarity), and fluency (perplexity score), description is given in Table~\ref{tab:metric_descriptions}. Each justification was assessed individually, and the resulting metric distributions are visualized in Figure~\ref{fig:justification_metrics_subfigs}.


\begin{table}[h]
\centering
\caption{Descriptions of evaluation metrics used to assess justification quality.}
\begin{tabular}{|p{3.5cm}|p{9cm}|}
\toprule
\hline
\textbf{Metric} & \textbf{Description} \\
\hline
\midrule
\textbf{Readability (Flesch Reading Ease)~\cite{fleschKincaidWiki}} & Measures ease of understanding based on sentence length and syllable count. Higher scores indicate more readable text. \\
\hline
\textbf{Lexical Richness (TTR)} & Type-Token Ratio calculates the ratio of unique words to total words in the justification. Higher TTR values indicate more varied vocabulary. \\
\hline
\textbf{Embedding Coherence} & Computes average cosine similarity between sentence embeddings using a pre-trained transformer. Higher values suggest better contextual and semantic flow. \\
\hline
\textbf{Perplexity score} & Measures fluency based on how well a generative model predicts the sequence. Lower values indicate more natural and fluent language. \\
\hline
\bottomrule
\end{tabular}
\label{tab:metric_descriptions}
\end{table}

The automatic evaluation of AURA’s generated justifications reveals encouraging results across multiple dimensions of textual quality. The average readability score was 27.28, which is consistent with the expected complexity of formal cyber threat intelligence reports. This level of readability indicates that the language is appropriately technical and tailored for professional analysts rather than general audiences. Lexical richness, measured through a Type-Token Ratio (TTR) of 0.67, reflects the use of diverse vocabulary, suggesting that the justifications are informative and avoid excessive repetition. Embedding coherence yielded an average cosine similarity of 0.33 between sentence embeddings, indicating moderate to strong semantic flow and contextual alignment across sentences. Finally, the average perplexity score of 57.05, while higher than general-domain benchmarks, remains acceptable for domain-specific narratives where specialized terminology and structured reasoning are common. These results collectively affirm that AURA’s justifications are not only technically sound but also linguistically coherent and suitable for expert interpretation.

\subsubsection{LLM-as-Judge Evaluation}
LLMs have been increasingly used as automated evaluators or "judges" for assessing the quality of generated content, offering scalable and consistent evaluations~\cite{gu2024survey,zheng2023judging}. To complement the automated metrics, we also employed a language model-based evaluation. Specifically, gpt-4o was prompted to act as an expert evaluator, scoring each justification on a 1–10 scale across four dimensions: fluency, clarity, coherence, and informativeness. The model was given the following prompt:
\begin{tcolorbox}[title={Prompt to LLM-as-Judge}, breakable, enhanced, colback=gray!5, colframe=black, fonttitle=\bfseries, boxrule=0.5pt]
\begin{verbatim}
You are an expert language evaluator. Rate the 
following paragraph on a scale of 1 to 10 for each 
of the following:
1. Fluency (grammar and flow)
2. Clarity (ease of understanding)
3. Coherence (logical structure and topic continuity)
4. Informativeness (useful and relevant information)

Paragraph:
"""<paragraph>"""

Return your answer as a JSON object:
{
  "fluency": number,
  "clarity": number,
  "coherence": number,
  "informativeness": number
}
\end{verbatim}
\end{tcolorbox}

\begin{figure}[h]
    \centering
    \includegraphics[width=0.9\linewidth]{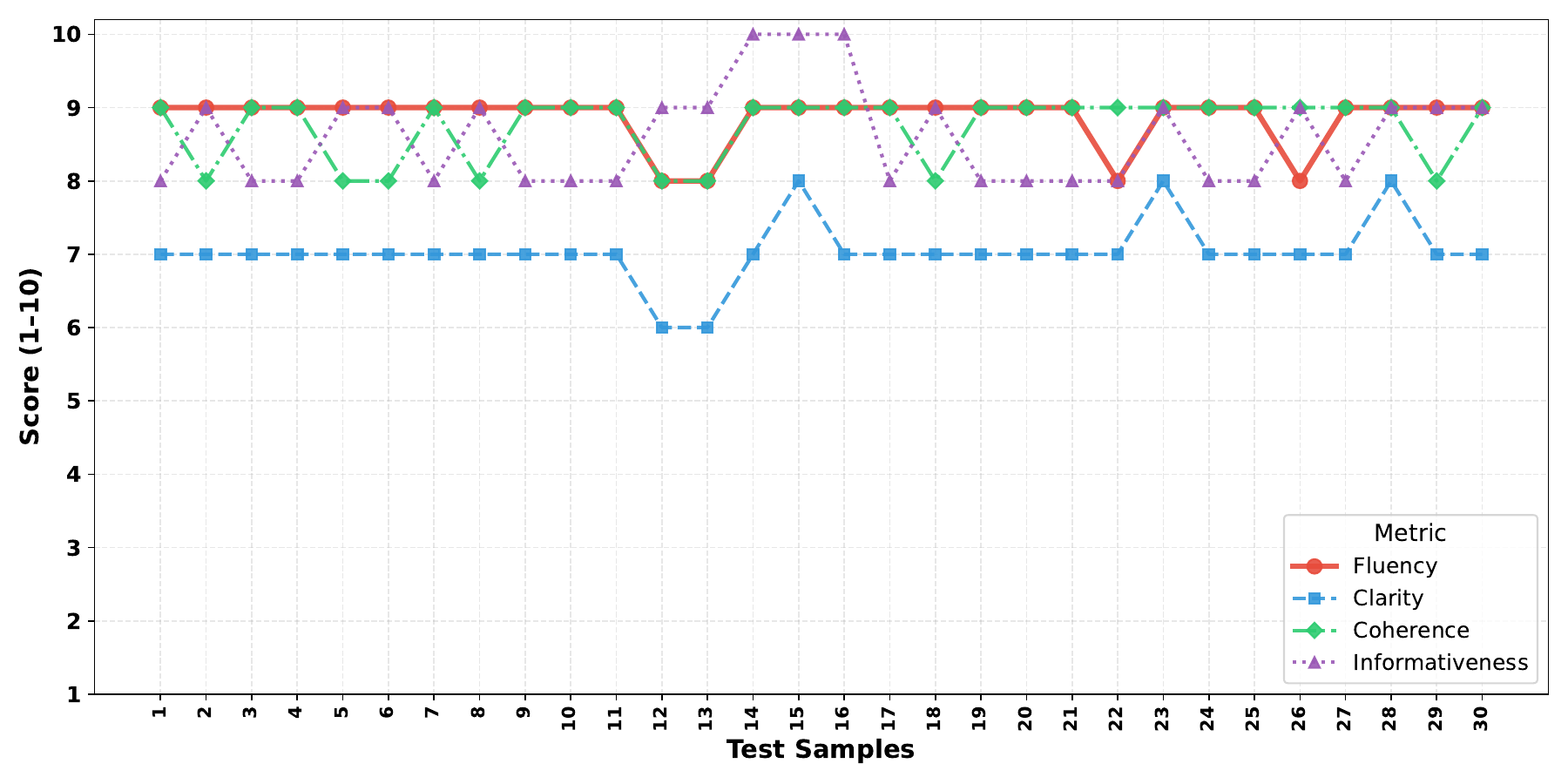}
    \caption{Combined LLM-as-Judge evaluation of justifications across four dimensions (1–10 scale). Each line represents a metric evaluated per justification. Fluency is emphasized for visual clarity.}
    \label{fig:llm_judge_merged}
\end{figure}

\noindent The average scores were notably high: 8.87 for fluency, 7.03 for clarity, 8.73 for coherence, and 8.6 for informativeness, indicating consistent linguistic and semantic quality. Figure~\ref{fig:llm_judge_merged} presents a consolidated view of these scores, highlighting consistent trends across justifications and test samples.

\noindent These two perspectives together affirm that AURA produces justifications that are not only grounded in threat intelligence evidence but are also linguistically fluent, semantically coherent, and informative—critical attributes that enhance analyst trust and support operational decision-making in cyber threat analysis workflows.

\subsection{Case Study}
\label{subsec:case_study}

To demonstrate  AURA’s real-world attribution capabilities, we present a case study derived from publicly available threat intelligence reports. This examples showcase AURA’s ability to perform both group-level and nation-level attribution by synthesizing technical indicators (e.g., TTPs, IOCs, tooling) with contextual signals (e.g., geography, targeting, and infrastructure). It also shows distinct analytical challenges, such as actor overlap and deceptive infrastructure, providing a realistic evaluation of AURA’s multiagent reasoning performance.

\subsubsection{APT36 – Youth Laptop Scheme Phishing Campaign}

This case study focuses on a cyber espionage campaign uncovered by Cyfirma during March 2025, where adversaries exploited Pakistan's youth laptop scheme as a lure to target sensitive Indian sectors. The campaign used decoy documents and spear-phishing tactics to compromise users across defense, aerospace, education, and government domains.

\paragraph{Threat Indicators}  
The operation featured a mix of phishing (\texttt{T1566}), malicious PowerShell execution (\texttt{T1059.001}), and encrypted communication channels (\texttt{T1573}). Tools deployed included \textit{Crimson RAT}, \textit{Poseidon}, and \textit{ElizaRAT}, all of which enabled remote access, clipboard monitoring, and location tracking (\texttt{T1115}, \texttt{T1430}). Infrastructure impersonated Indian government themes (e.g., \texttt{email.gov.in.gov-in.mywire.org}) to enhance social engineering effectiveness. A summary of the observed threat artifacts is presented in Table~\ref{tab:youthscheme_artifacts}.

\begin{table}[H]
\centering
\caption{Extracted Threat Artifacts from the Youth Laptop Scheme Campaign}
\label{tab:youthscheme_artifacts}
\begin{tabular}{|l|p{9cm}|}
\hline
\textbf{Malware/Tools} & Crimson RAT, ElizaRAT, Poseidon \\
\hline
\textbf{Key TTPs} & 
T1059.001 (PowerShell), T1071 (Web Protocols), T1115 (Clipboard Capture), T1204 (User Execution), 
T1409 (Stored App Data), T1430 (Location Tracking), T1546.013 (PowerShell Profile), 
T1566 (Phishing), T1573 (Encrypted Channel) \\
\hline
\textbf{IOCs} & 
88[.]222[.]245[.]211,\newline email[.]gov[.]in[.]gov-in[.]mywire[.]org,\newline postindia[.]site,\newline 
287a5f95458301c632d6aa02de26d7fd9b63c6661af33\newline1dff1e9b2264d150d23, cbf74574278a22f1c38ca922f91548596630fc67bb2348\newline34d52557371b9abf5d \\
\hline
\textbf{Targets} & India, Government agencies, Aerospace, Defense contractors, Educational institutions, Military \\
\hline
\textbf{Campaign Timeline} & Active during 2024–2025 \\
\hline
\end{tabular}
\end{table}

\paragraph{AURA Attribution Output}  
AURA attributed this campaign to \textit{APT36} (Transparent Tribe) as the primary actor, with \textit{APT37} listed as a secondary candidate. Group-level attribution was driven by the use of Crimson RAT, themed phishing infrastructure, and PowerShell-based payloads—hallmarks of APT36 operations. Nation-level attribution points to \textit{Pakistan}, as APT36 is consistently associated with Pakistan-based interests targeting India.

\begin{tcolorbox}[title=AURA Attribution Justification, breakable, enhanced, colback=gray!5, colframe=black, fonttitle=\bfseries, boxrule=0.5pt]
APT36, also known as Transparent Tribe, has a history of targeting government and military organizations in India... The use of Indian-themed infrastructure such as email[.]gov[.]in[.]gov-in[.]mywire[.]org and postindia.site further aligns with APT36’s known campaigns.
\end{tcolorbox}

\paragraph{Analytical Insight}  
APT36 has long exploited geopolitical narratives for social engineering, including student- or education-themed phishing lures. The use of Crimson RAT, India-themed infrastructure, and educational targeting strengthens attribution confidence. Public threat intelligence by Cyfirma~\cite{cyfirma2024youthscheme} directly links this campaign to APT36. Furthermore, MITRE ATT\&CK documentation~\cite{mitreattackapt36} confirms the group’s historical use of phishing, PowerShell exploitation, and targeting of Indian defense and government entities. APT37 was considered a secondary actor due to partial TTP overlap (e.g., encrypted channels and PowerShell), but lacks historical focus on Indian targets. This case reinforces AURA’s effectiveness in leveraging contextual indicators, particularly infrastructure and thematic deception, for attributing threats to nation-linked actors.
\\
\\
This case study shows that AURA effectively handles the complexity of real-world attribution. It demonstrates reliable, interpretable, and actor-aligned outputs across diverse campaigns. The system’s integration of structured knowledge, contextual retrieval, and natural language justification makes it a valuable decision-support tool for cyber threat analysts.

\subsection{Representative Examples of Overlapping Attribution}
AURA exhibits robustness in complex attribution settings where behavioral indicators are ambiguous or shared among multiple threat groups. This is especially evident in cases where the groups originate from the same country. This is reflected in the consistent gains observed when evaluating top-2 accuracy instead of top-1, showing that AURA often captures the correct actor even when the top prediction is a closely related peer group.


Such overlaps are common in real-world cyber operations, where nation-state-sponsored APT groups may share tooling, infrastructure, or strategic objectives. Table~\ref{tab:ambiguity-cases} summarizes two illustrative examples from our evaluation.

\begin{table}[h]
\centering
\caption{Top predictions for ambiguous attribution cases.}
\label{tab:ambiguity-cases}
\begin{tabular}{|p{4cm}|p{2.5cm}|p{2.5cm}|p{2.5cm}|}
\hline
\textbf{Report File} & \textbf{True Actor} & \textbf{Top-1 Pred.} & \textbf{Top-2 Pred.} \\
\hline
Trend Micro’s threat report on Pawn Storm~\cite{trendmicro2024pawnstorm} & APT28 & APT29 & APT28 \\
\hline
Zscaler’s threat report on the Lazarus Group~\cite{zscaler2024lazarusnaver} & Lazarus Group & Kimsuky & Lazarus Group \\
\hline
\end{tabular}
\end{table}

\begin{tcolorbox}[title=Justification for Trend Micro's Threat Analysis, breakable, enhanced, colback=gray!5, colframe=black, fonttitle=\bfseries, boxrule=0.5pt]
APT29 is known for targeting government, military, and energy sectors, aligning with the observed targets like Defense, Embassies, and Governments. Their typical tactics include T1041: Exfiltration Over C2 Channel, T1078: Valid Accounts, and sophisticated phishing (T1566). They have historically used application layer protocols for communication (T1071) and are linked to similar geographic regions like Europe and North America. APT28, also known as Fancy Bear, is another possibility due to its known use of Tactics, Techniques, and Procedures (TTPs) such as use of Valid Accounts (T1078) and Brute Force (T1110), and targeting of governmental and military entities. They have also been historically associated with targeting in Middle East, Europe, and North America. Both actors have shown capability and intent to operate across the specified regions and sectors.
\end{tcolorbox}

\begin{tcolorbox}[title=Justification for Zscaler's Threat Analysis, breakable, enhanced, colback=gray!5, colframe=black, fonttitle=\bfseries, boxrule=0.5pt]
The primary actor Kimsuky is known for targeting South Korea and using spear phishing (T1566) and application layer protocol (T1071) techniques. They have historically focused on South Korean entities. Moreover, the inclusion of South Korea as a target and use of email domains similar to Naver, a popular South Korean email provider, aligns with Kimsuky's modus operandi. The secondary actor Lazarus Group is also possible given their involvement in cybercrime activities including cryptocurrency theft. Lazarus has a history of targeting South Korea and using similar TTPs. However, Kimsuky is ranked primary due to the specific targeting associated with South Korean domains and mail services.
\end{tcolorbox}

In the first case, AURA predicted \textit{APT29} as the primary actor and \textit{APT28} as the secondary, while the ground truth was APT28. The justification notes APT29’s alignment with targets such as defense and embassies using TTPs like \texttt{T1041}, \texttt{T1078}, and \texttt{T1566}, while also acknowledging that APT28, known for \texttt{T1078} and \texttt{T1110}, is a credible candidate due to similar geopolitical targeting. Both actors, affiliated with Russia, have historically operated across Europe and North America, often blurring attribution boundaries.

In the second case, AURA ranked \textit{Kimsuky} higher than the ground truth \textit{Lazarus Group}. The justification emphasizes Kimsuky's use of spearphishing (\texttt{T1566}) and infrastructure mimicking South Korean email services, particularly Naver-like domains, which closely matched the campaign’s characteristics. Although Lazarus Group was also identified as a plausible actor due to its overlapping TTPs and history of targeting South Korea, Kimsuky was favored because of its stronger alignment with domain-specific artifacts. Both groups are North Korea-affiliated and share similar targeting patterns, making attribution inherently ambiguous in such scenarios.

These examples emphasize AURA's robustness: even when ambiguity arises from overlapping modus operandi, AURA surfaces both likely actors, enabling analysts to consider high-confidence alternatives within the same geopolitical context. Rather than misattributing entirely, AURA reflects the behavioral convergence between actors, reinforcing its practical utility in real-world, multi-campaign threat intelligence workflows.
\\
\\
\noindent Overall, the findings suggest that the AURA framework demonstrates competitive accuracy while also incorporating key human-centric elements such as modularity, interpretability, and robustness. These characteristics make it well-suited for operational use in threat intelligence and attribution workflows.

\section{Limitations and Future Directions}
\label{sec:AuraLimitationAndFutureWork}

While AURA demonstrates strong performance, several limitations remain. First, the current evaluation uses a relatively small test set comprising $30$ threat reports. The limited test size results from our effort to exclude samples that may have been part of LLM training, ensuring an unbiased evaluation based on post-cutoff threat reports. Although sufficient for controlled analysis, a larger and more diverse evaluation set is essential for broader generalizability. In future work, we plan to scale the evaluation using an expanded testbed that includes both public and proprietary datasets to validate AURA across varied recent threat scenarios.

Further, the Justification Agent currently provides textual explanations without evidence weighting. Future versions may benefit from incorporating explicit reasoning chains or confidence scoring to support analyst decision-making.

We also plan to evaluate the AURA framework using open-source LLMs to promote transparent and reproducible evaluation of intelligent systems. A further extension is to expand AURA's capabilities to support more granular levels of attribution, including intrusion set–level and campaign–level linking, as well as to assess its performance on multilingual threat reports and streaming data.


\section{Conclusion}
\label{sec:AuraConclusion}

In this work, we introduced AURA (Attribution Using Retrieval-Augmented Agents), a retrieval-augmented, multi-agent framework for cyber threat attribution that combines the reasoning capabilities of LLMs with structured threat intelligence. AURA demonstrated strong performance across group-wise and nation-wise attribution tasks, producing accurate and interpretable results supported by contextual justifications. Our experiments with black-box LLMs validate the effectiveness of modular agent design and evidence-guided reasoning. While the current evaluation is constrained by dataset size, AURA establishes a strong foundation for scalable and transparent attribution workflows. Future directions include the use of larger datasets, more granular attribution, and deeper justification of attribution decisions.

\bibliographystyle{acm} 
\bibliography{cas-refs}

\end{document}